\documentclass[ twocolumn]{aastex631}

\usepackage{makecell}   

\def\nustar{{\it NuSTAR}}
\def\integral{{\it INTEGRAL}}
\def\integralib{{\it INTEGRAL-IBIS}}

\def\swift{{\it Swift}}

\def\maxi{{\it MAXI}}

\def\straycats{{\tt StrayCats}}

\def \caltech {{Cahill Center for Astronomy and Astrophysics, California Institute of Technology, Pasadena, CA 91125, USA}}

\shorttitle{NuSTAR beyond 79 keV}
\shortauthors{Mastroserio et al.}
\graphicspath{{./}{figures/}}

\begin{document}
\title{NuSTAR spectral analysis beyond 79 keV with stray light}

\author[0000-0003-4216-7936]{G.~Mastroserio}
\affiliation{\caltech{}}
\author[0000-0002-1984-2932]{B.~W.~Grefenstette}
\affiliation{\caltech{}}
\author[0000-0001-6269-2821]{P.~Thalhammer}
\affiliation{Dr. Karl Remeis-Observatory, University of Erlangen-Nuremberg, Sternwartstr. 7, 96049 Bamberg, Germany}
\author[0000-0002-5341-6929]{D.~J.~K.~Buisson}
\affiliation{unaffiliated}
\author[0000-0002-4024-6967]{M.~C.~Brumback}
\affiliation{\caltech{}}
\author[0000-0002-8961-939X]{R.~M.~Ludlam}\thanks{NASA Einstein Fellow}
\affiliation{\caltech{}}
\affiliation{Department of Physics $\&$ Astronomy, Wayne State University, 666 West Hancock Street, Detroit, MI 48201, USA}
\author[0000-0002-8908-759X]{R.~M.~T.~Connors}
\affiliation{\caltech{}}
\author[0000-0003-3828-2448]{J.~A.~Garc\'ia}
\affiliation{\caltech{}}
\author[0000-0003-2538-0188]{V. Grinberg}
\affiliation{European Space Agency (ESA), European Space Research and Technology Centre (ESTEC), Keplerlaan 1, 2201 AZ Noordwik, the 15 Netherlands}
\author[0000-0003-1252-4891]{K.~K.~Madsen}
\affiliation{CRESST and X-ray Astrophysics Laboratory, NASA Goddard Space Flight Center, Greenbelt, MD 20771 USA}
\author{H.~Miyasaka}
\affiliation{\caltech{}}
\author[0000-0001-5506-9855]{J.~A.~Tomsick}
\affiliation{Space Sciences Laboratory, 7 Gauss Way, University of California, Berkeley, CA 94720-7450, USA}
\author{J. Wilms}
\affiliation{Dr. Karl Remeis-Observatory, University of Erlangen-Nuremberg, Sternwartstr. 7, 96049 Bamberg, Germany}



\begin{abstract}
Due to the structure of the \nustar\ telescope, photons at large off-axis 
($> 1^{\circ}$) can reach the detectors directly (stray light), without passing through the instrument optics. 
At these off-axis angles \nustar\ essentially turns into a collimated instrument and the 
spectrum can extend to energies above the Pt k-edge (79~keV) of the multi-layers, which limits the 
effective area bandpass of the optics. 
We present the first scientific spectral analysis beyond 79~keV 
using a Cygnus~X-1 observation in \straycats, the catalog of stray light observations.
This serendipitous stray light observation occurred simultaneously with an \integral\ observation.
When the spectra are modeled together in the 30-120~keV energy band,  
we find that the \nustar\ stray light flux is well calibrated and 
constrained to be consistent with the \integral\ flux at the 90$\%$ confidence level.
Furthermore, we explain how to treat the background of the stray light spectral analysis, 
which is especially important at high energies. 
\end{abstract}


\section{Introduction} \label{sec:intro}
The X-ray detectors on board of the Nuclear Spectroscopic Telescope Array (\nustar; \citealt{Harrison2013})
are able to detect photons in the energy range between $2$-$160$ keV. 
Due to the separation between the optics and the detectors and the Pt/C multi-layers deposited on the mirrors, 
\nustar\ is able to focus hard X-ray photons in the 3-79~keV bandpass. 
However, because of the open geometry of the mast that connects the 
optics and the focal plane modules (FPMs), photons can also reach the detectors without 
passing through the optics. 

Sources within $1^{\circ}$-$5^{\circ}$ of the primary target cause stray light, 
also referred as 'aperture flux', and when bright enough will 
leave a distinct circular pattern \citep{Madsen2017b}. 
This pattern is easy to predict and identify on the detectors, 
especially when produced by a single source.
Although the position angle of the observatory is typically optimized to minimize 
the stray light pattern for a given observation  
(since it may cause additional background for the focused target), 
in some cases it is unavoidable (\citealt{Grefenstette2021}). 
These serendipitous observations of sources that are not the focused target 
can be of great use to perform additional science for several reasons:  
1) they can be used to monitor a source even though a focused observation was not scheduled; and
2) they can be used to expand the \nustar\ energy range above 79~keV since the
 stray light does not go through the optics where the Pt K-edge of the multi-layers 
 coatings drastically decreases the response of the instrument.

\citet{Madsen2017a} considered stay light observations of the Crab during 2 different epochs and 
they showed that it is possible to reconstruct 
the response for stray light sources, and to perform spectral analysis 
of X-ray sources. 
\citet{Grefenstette2021} presented the first stray light catalog (\straycats) wherein they examined 1400
\nustar\ observations up to mid 2020 and identified nearly 80 stray light sources. 
The \straycats\ has recently been updated \citep[ \straycats\  v2][]{Ludlam2022b} 
to include additional \nustar\ observations up to the beginning of 2022. 
Moreover, \straycats\ v2 contains new information to characterise the behaviour of the sources
such as long term \maxi\ and \swift\ light curves of the stray light sources, 
region files to extract products, mean intensity for each stray light detection, 
and a proxy for the background level based on the `empty' detector region 
(excluding the stray light region and the primary source).
\citet{Brumback2022} reported the first scientific timing analysis of a stray light source. 
They took advantage of the serendipitous stray light observations of SMC~X-1 (a high mass X-ray binary) 
to confirm the orbital ephemeris and study the shape of the pulse profile over time. 
They show that stray light observations are suitable to perform timing analyses of X-ray binaries. 

In this work, we consider Cygnus~X-1, a well studied high mass X-ray binary 
hosting a stellar mass black hole. 
This source is a great candidate to extend the high energy limit of the \nustar\ spectrum 
since it is one of the brightest persistent X-ray binary in the sky
that exhibits a hard X-ray spectrum in the sky \citep[e.g.][]{DelSanto2013,Grinberg2013,Walter2017,Cangemi2021}. 
We perform the first spectral analysis of a stray light observation and 
demonstrate that it is possible to extend spectral analysis above 79~keV.
We compare the stray light energy spectrum with the \integralib\ spectrum \citep{Ubertini2003,Lebrun2003}.
The \nustar\ flux at high energies (above 30 keV) of the stray light is equivalent to the
the \integralib\ flux within 90$\%$ confidence level, 
thus demonstrating that the nominal \nustar\ energy band can be extended to
higher energies when properly accounting for the background.

\section{\nustar\ stray light data} 
\label{sec:data_reduction_NuSTAR}

\begin{table*}
\caption{Some characteristics of the \nustar\ stray light observation analyzed in this work.}
\begin{tabular}{ c|c|c|c|c|c }
  \hline
ObsID & Date &SL source & Primary & Exposure & Area SL \\

  \hline
90501328002 & 2019 June 4 & Cygnus X-1  & 4U 1954+31  & 40791 [s]& 0.82 cm$^2$ \\
  \hline

  \end{tabular}

    \label{tab:summary}
 \end{table*}

\begin{figure}
	\includegraphics[width=\columnwidth]{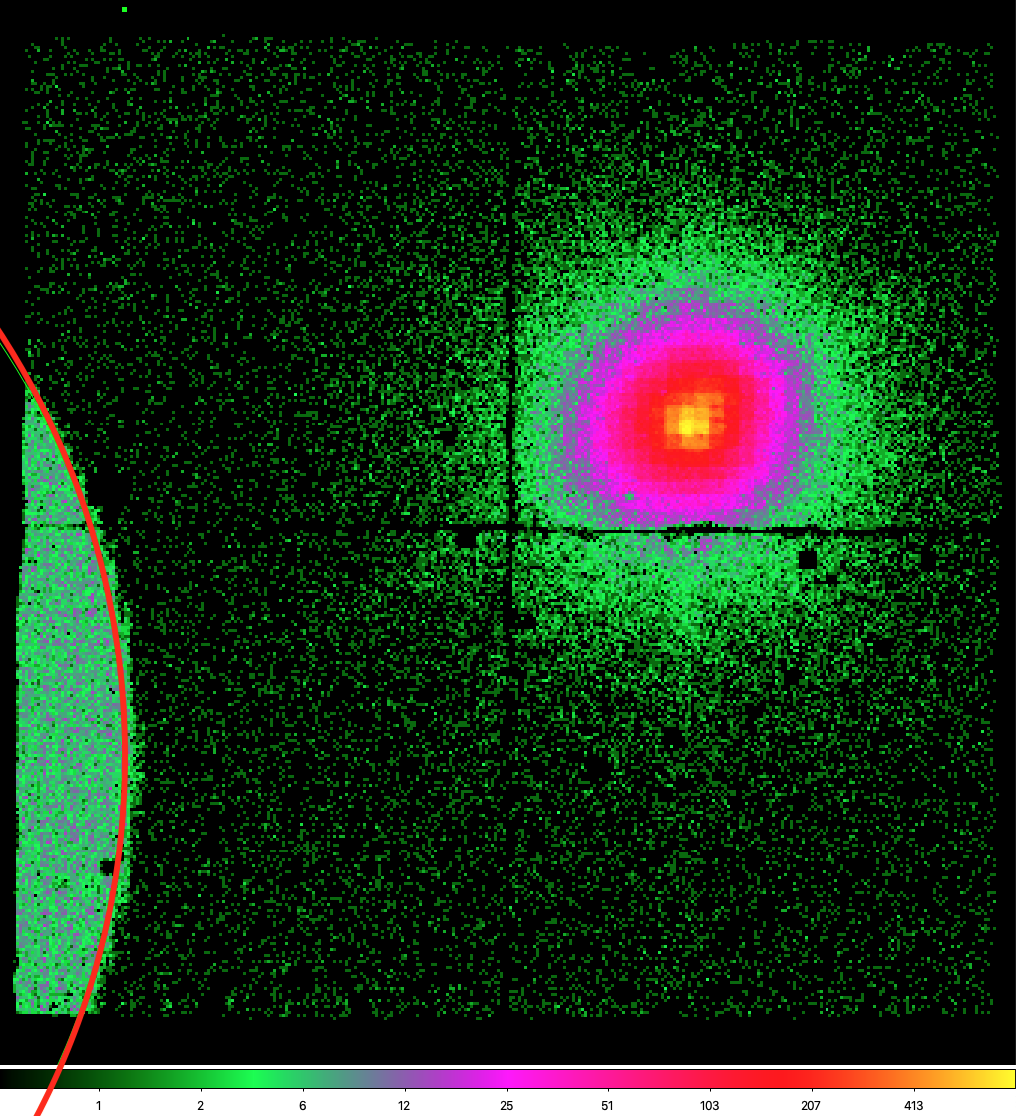}
    \caption{Image of the \nustar\ FPMA in Det$\#1$ coordinates. The red thick line indicates the stray light region. The primary source (4U~1954+31) is visible in Det$\#0$ (upper right quadrant). }
    \label{fig:straylight_region}
\end{figure}

Cygnus X-1 was observed 3 times in stray light prior to 2021. 
We consider observation 90501328002, since this is the only observation with simultaneous \integral\ exposure.  
Table~\ref{tab:summary} provides information about the observation. 
Stray light is visible on Det$\#2$ and marginally on Det$\#1$ of FPMA as shown in Fig~\ref{fig:straylight_region}. 
The FPMB detectors do not show evidence of stray light contamination, thus no analysis of FPMB is needed. 
The primary source (4U~1954+31) is positioned in Det$\#0$ and affects a portion of 
the field of view separate from the stray light region, 
thus it is easy to exclude during the analysis. 
We use \texttt{SAOImage DS9} to define the circular segment region where the stray light is present 
(see thick red line in Fig~\ref{fig:straylight_region}). 
We define the stray light region using a circular region with a radius of $500$ arcseconds and 
centered outside the field of view since that area does not contribute to the event file. 

\begin{figure}
	\includegraphics[width=\columnwidth]{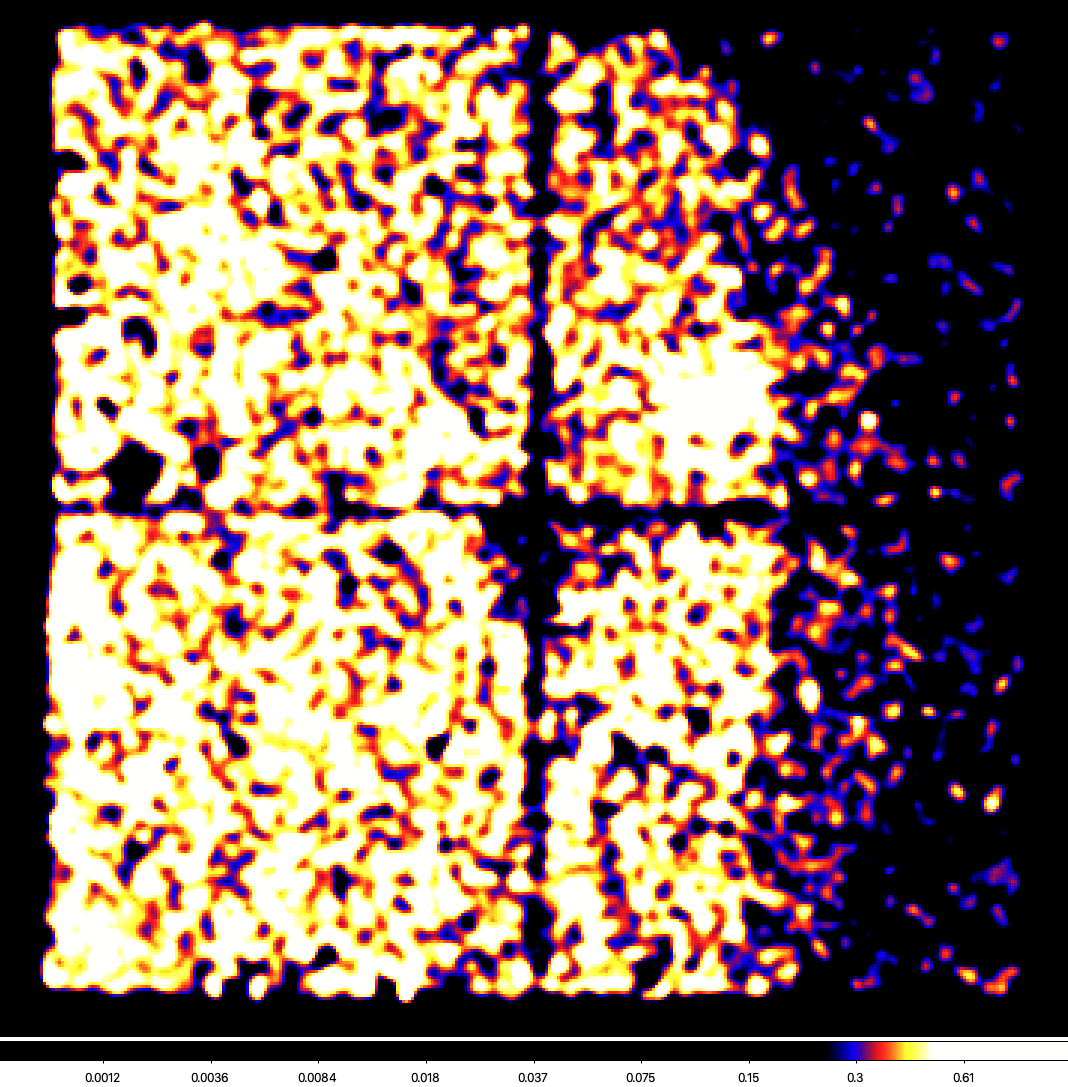}
    \caption{Image of \nustar\ FPMA for energies $> 50$ keV in Det$\#1$ coordinates. A smoothing Gaussian function has been applied. The stray light contaminates all detectors. }
    \label{fig:straylight_smoothing}
\end{figure}

The background extraction is problematic for this observation. 
The aperture stops in front of the detectors are designed to block the low energy stray light photons producing 
the characteristic stray light pattern in Fig.~\ref{fig:straylight_region}.
However, the materials of aperture stops become transparent to the high energy photons above 50~keV \citep[][]{Madsen2017b}. 
Fig.~\ref{fig:straylight_smoothing} shows the \nustar\ field of view filtered 
for energies above 50~keV that has been smoothed with a Gaussian function (radius $ = 5$ and  $\sigma = 2$). 
The Cygnus~X-1 stray light contaminates all 4 detectors which makes it 
impossible to select a background region valid for the high energies.

\citet{Wik2014} showed that above 30~keV the dominant background component is 
the internal background of the instrument which consists of the internal continuum 
and emission lines (see \texttt{nuskybgd}\footnote{www.github.com/NuSTAR/nuskybgd} and Fig.~10 in \citealt{Wik2014}). 
For the purpose of this work we can neglect the contribution of the internal lines 
and focus on correctly estimating the internal continuum background that dominates over the internal lines above 120~keV.
Therefore, when we fit the energy spectrum we consider up to 160~keV, 
and we use a constant background model with free normalization 
that is mainly constrained by the counts in the 120-160~keV range. 

We produce stray light science products (including the flux-energy spectrum) using the StraylightWrapperExample notebook in the \texttt{nustar-gen-utils}\footnote{https://github.com/NuSTAR/nustar-gen-utils}.
In particular, we use the functions `make$\_$det1$\_$image' to create the image, 
`extract$\_$det1$\_$events' to extract the event file, and 
`make$\_$det1$\_$spectra' together with `make$\_$exposure$\_$map', 
`make$\_$straylight$\_$arf' and `straylight$\_$area' to create the energy spectrum 
and the ARF response accounting for the correct area of the stray light. 
The response generation tools for stray light are consistent in flux normalisation with the nominal configuration of the CALBD.

\section{\integral\ data}
\integral\ observed Cygnus~X-1 in 2019 from May 24 until June 5. 
We select the portion of this observation simultaneous with  \nustar\ 
 to obtain the same exposure ($\sim$40 ks). 
For data reduction we use the Off-line Scientific Analysis (\texttt{OSA}) tools version 11.1. 
For this work we only make use of data from the \textit{ISGRI}, 
the upper detector layer of the \textit{IBIS} instruments. 
We generate a spectrum on a logarithmic energy grid from 25~keV to 400~keV with 30 bins, but
only consider energies above 30~keV during the spectral analysis due to the low energy threshold of \textit{ISGRI}.

\section{Spectral Analysis} 
\label{sec:analysis}

\begin{figure}
	\includegraphics[width=\columnwidth]{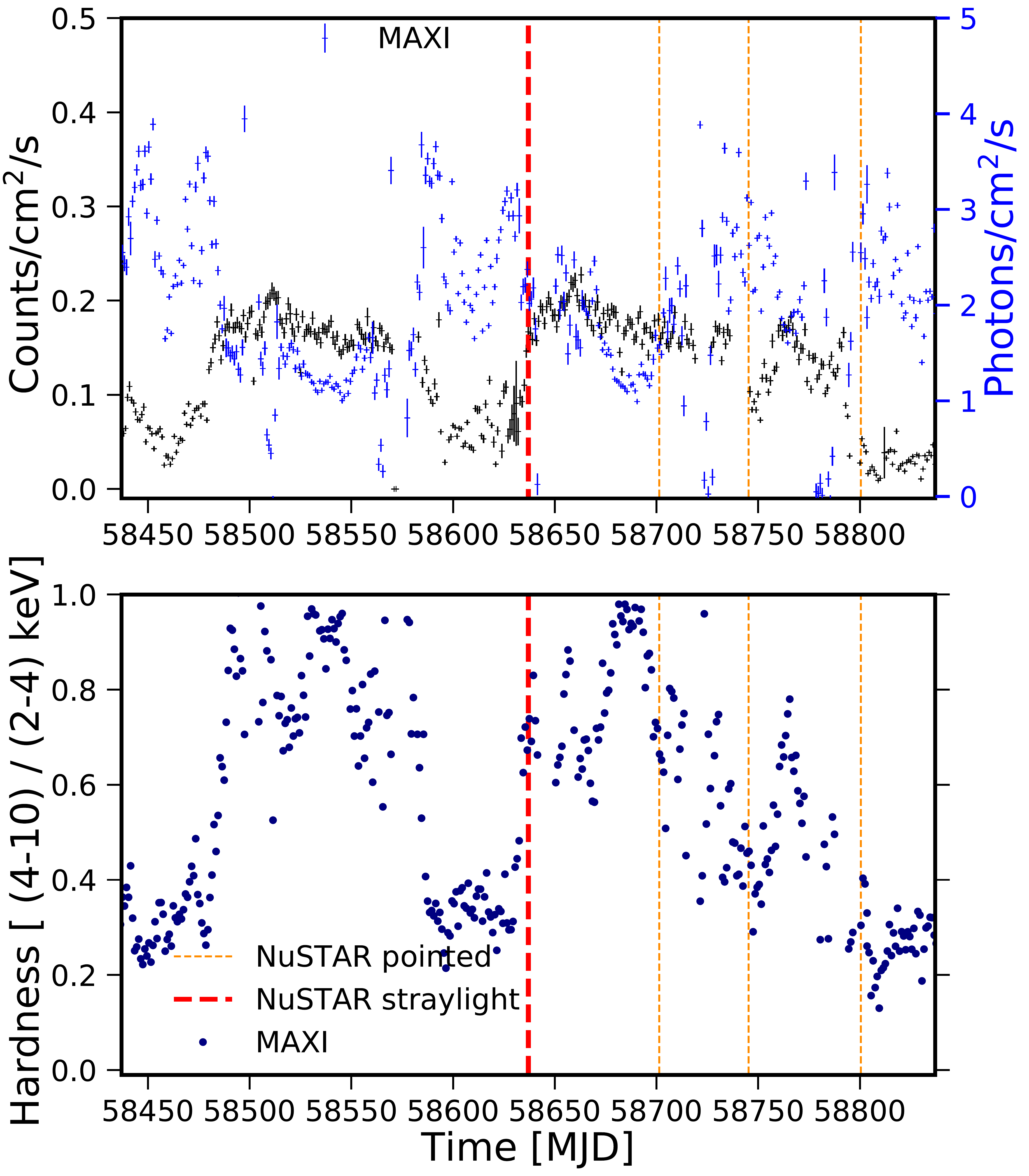}
    \caption{Top panel: \swift\ (black) and \maxi\ (blue) long term daily light curves of Cygnus~X-1. Bottom panel: hardness ratio as a 
    function of time of Cygnus~X-1 computed using \maxi\ energy bands. In both panels the red dashed lines 
    indicate when the stray light observation was performed and the orange lines indicate \nustar\ pointed observations of Cygnus~X-1.}
    \label{fig:lc_hardness}
\end{figure}

\begin{figure}
	\includegraphics[width=\columnwidth]{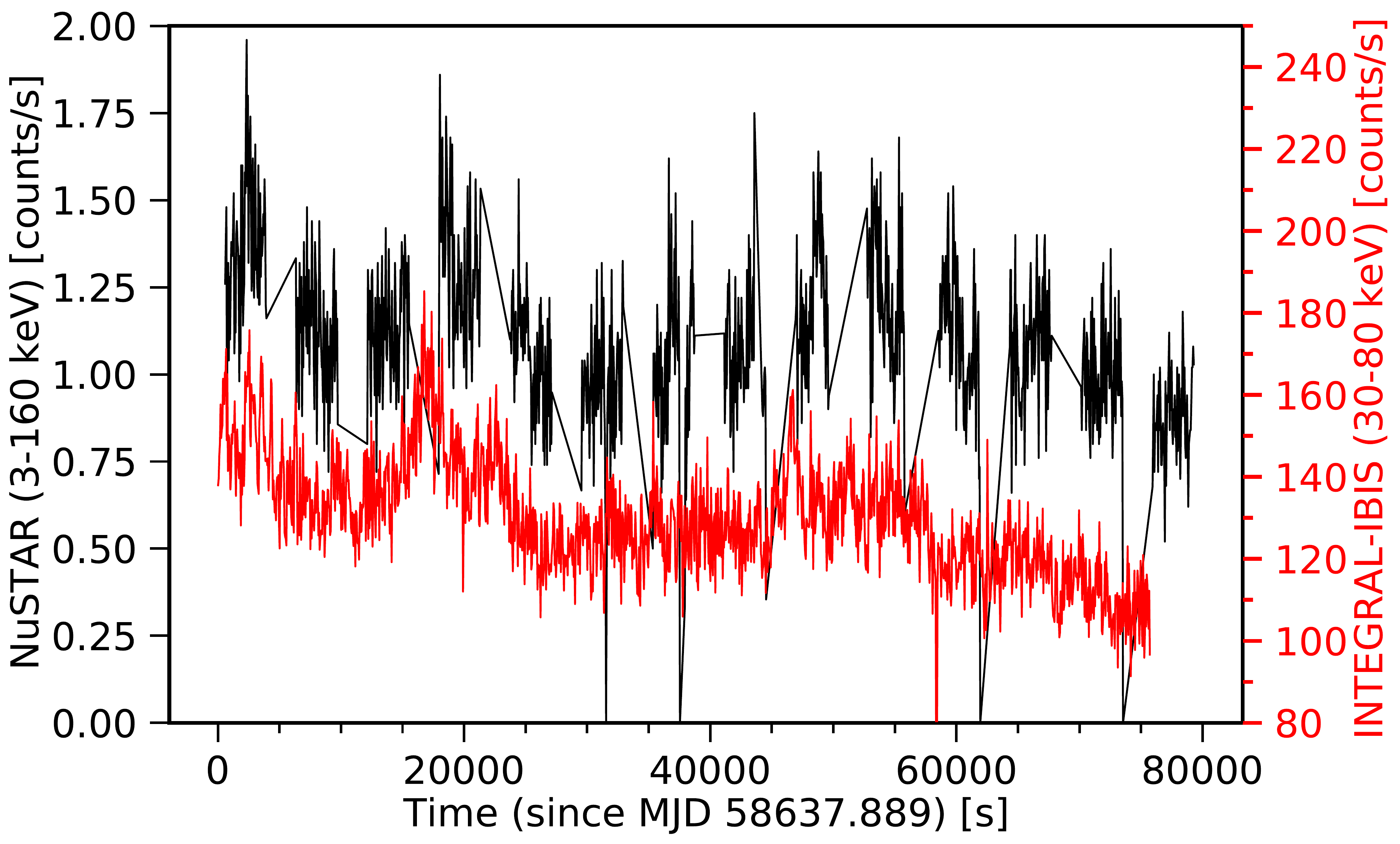}
    \caption{\nustar\ (black) and \integral\ (red) light curves during the simultaneous observation. }
    \label{fig:light_curves}
\end{figure}

\begin{figure}
	\includegraphics[width=\columnwidth]{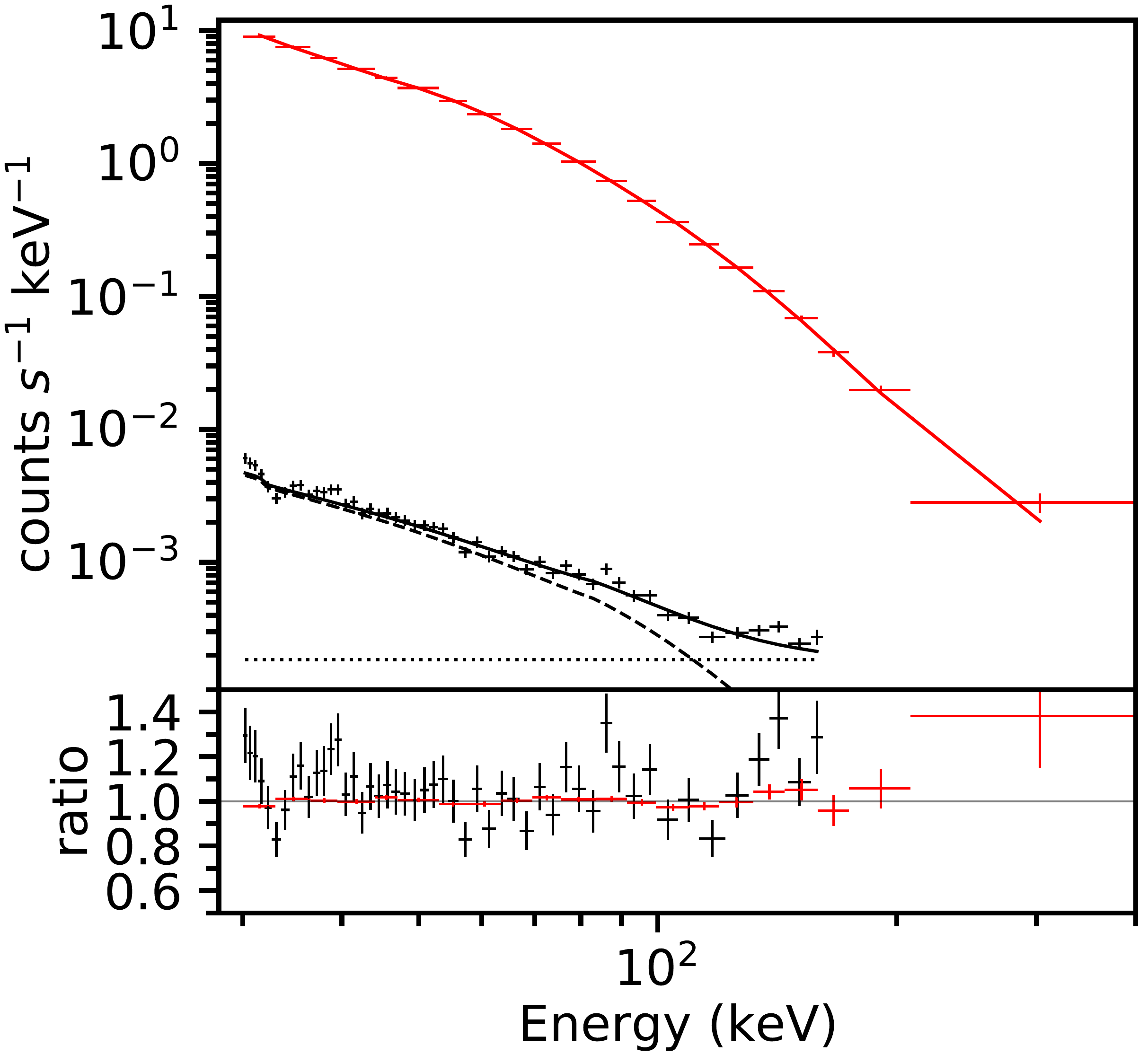}
    \caption{Best fit of the \nustar\ (black) and \integralib\ (red) flux energy spectrum above 30~keV with Model 1 (\texttt{cutoffpl}). 
    Top panel: Flux energy spectrum in units of counts per seconds per keV, both data and model. For the \nustar\ spectrum the dashed line is the cut-off power-law model, the dotted line is the background model, and the solid black line is the total model. Bottom panel: Residuals of the best fit.}
    \label{fig:pl_highEn}
\end{figure}

\begin{figure}
	\includegraphics[width=\columnwidth]{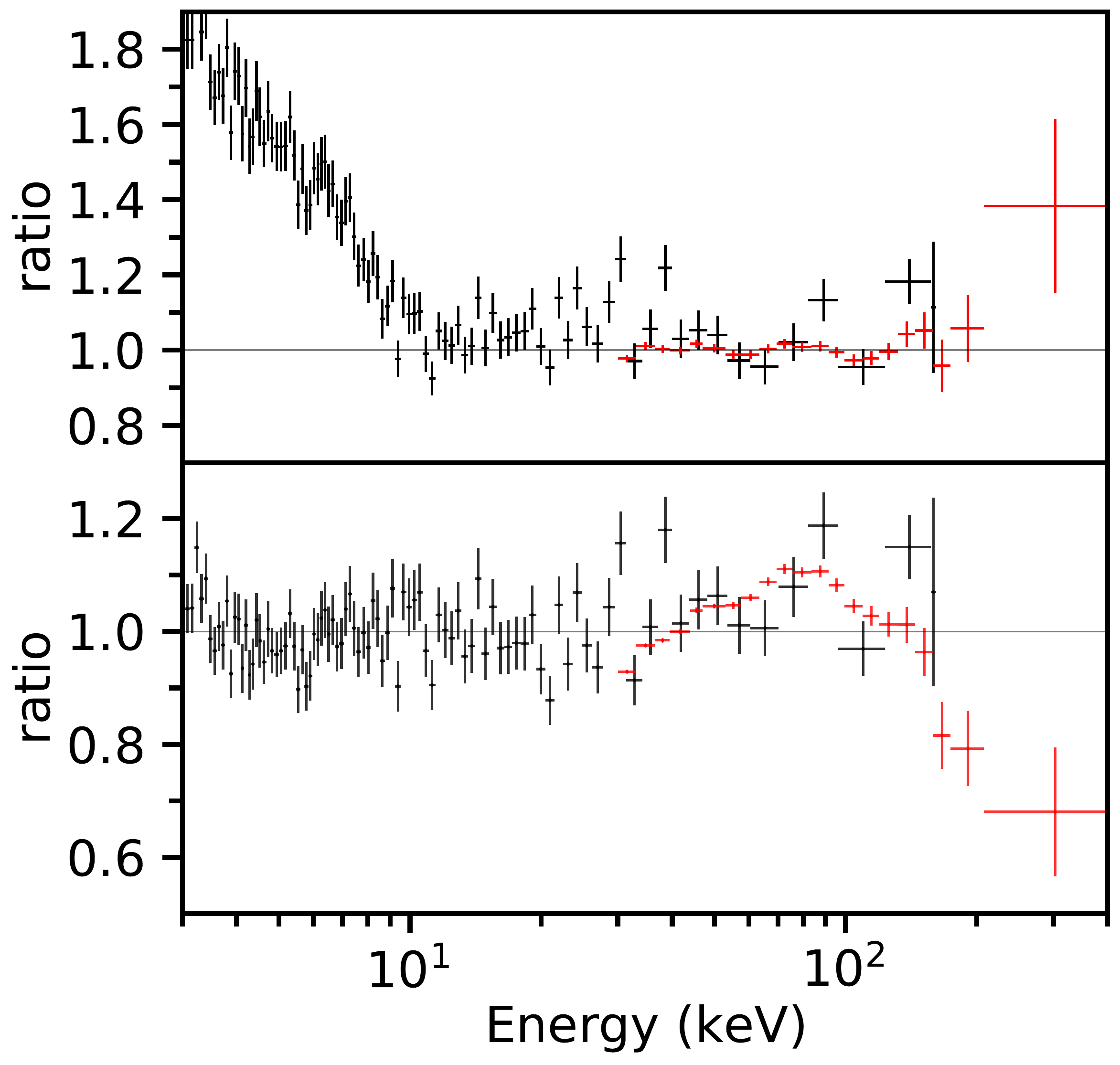}
    \caption{Top panel: residuals of the Model~1 (fit to the overlapping energy range) where we noticed the low energy channels of \nustar.  Bottom panel: residuals Model~2 (fit to the \nustar\ spectrum) where we added the \integralib\ spectrum and re-fit \textit{only} for the cross-calibration constant. All the model parameters are tied between \nustar\ and \integralib\ spectra.}
    \label{fig:residuals_gamma_fixed}
\end{figure}

\begin{figure}
	\includegraphics[width=\columnwidth]{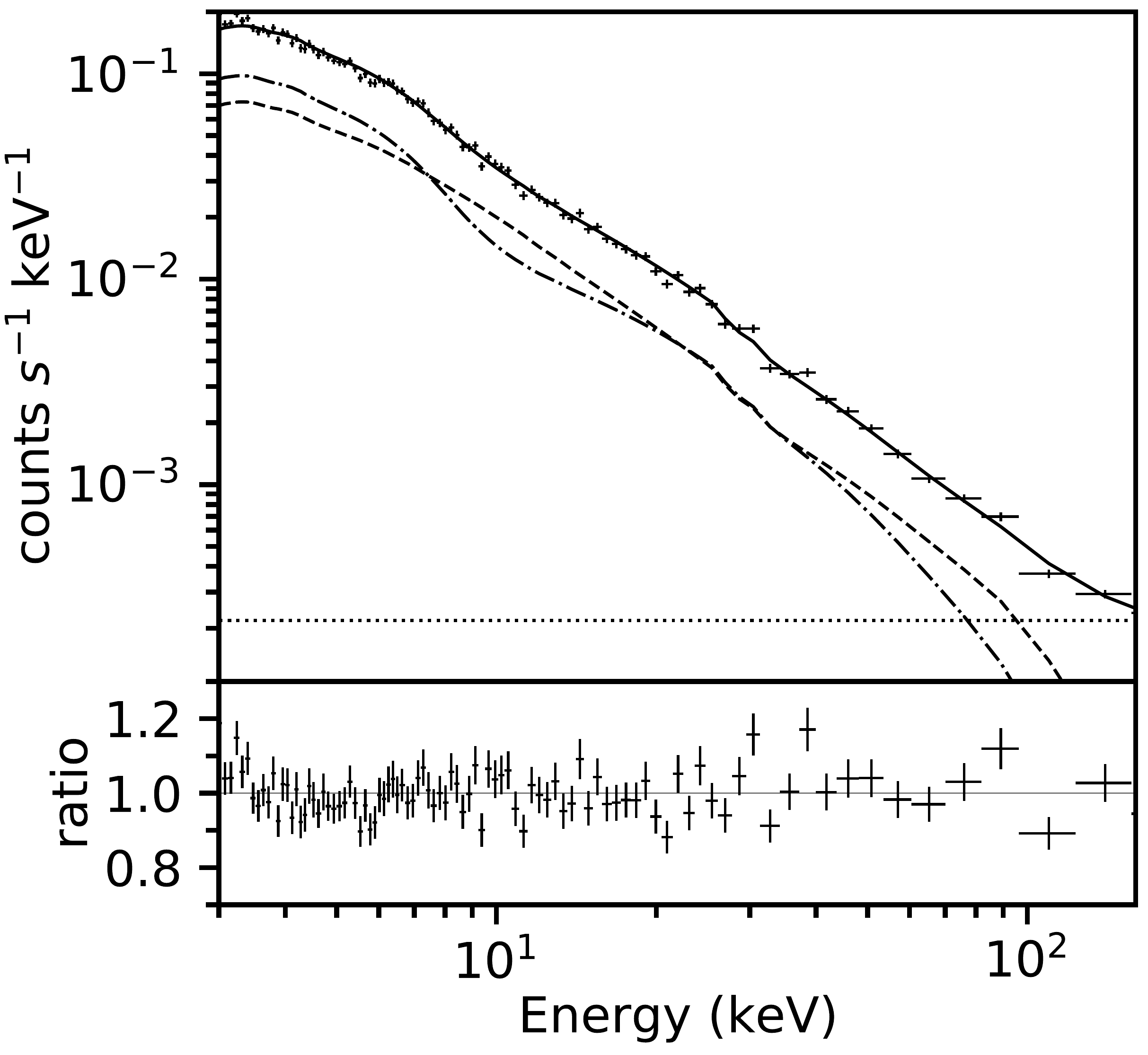}
    \caption{Best fit of the \nustar\ energy spectrum with Model~2 (\texttt{TBabs*[cutoffpl + relxill]}). The hydrogen column density is fixed to $6\times10^{21}$~cm$^{-2}$. Top panel: Flux energy spectrum in units of counts per seconds per keV, both data and model. The solid line is the total model. The non-solid lines are the additive components of the model: the dotted line is the background model, the dashed line is the continuum and the dashdotted line is the reflection. Bottom panel: Residuals of the best fit.}
    \label{fig:fit_only_nustar}
\end{figure}

\begin{table*}
\caption{Parameter values of the joint \nustar\ and \integral\ best fit. 
Model 1 (\texttt{cutoffpl}) considers only energy above $30$ keV for both instruments. 
For the rest of the models the energy range of \nustar\ extends down to $3$ keV.  
Model~2 is \texttt{TBabs*(cutoffpl + relxill)}. 
In the first fit we report the value of the reduced $\chi^2$ for comparison with the other 
best fits even though the fit is performed with Cash statistic. The flux in the overlapping energy band (30-120~keV) is 
the physical model flux for both \nustar\ (Nu) and \integral\ (INT). 
The column density is always fixed to $6\times10^{21}$~cm$^{-2}$. Statistical errors are quoted at $90\%$ confidence level.}
\noindent\begin{tabular}{@{} m{5.0cm}|m{2.5cm}|m{2.5cm}|m{2.5cm}|m{2.5cm}@{} }
  \hline
Parameters & Model 1 \newline (\footnotesize{Cash})$^*$ &  Model 1 \newline (\footnotesize{$\chi^2$}) & Model 2 \footnotesize{($\chi^2$, \nustar\ only)}&  
Model 2b \footnotesize{($\chi^2$)} \newline \footnotesize{(\nustar\ } + \newline \footnotesize{\textit{INTEGRAL})} \\

  \hline
Cross-cal           & $1.00^{+0.02}_{-0.01}$  &$1.02^{+0.03}_{-0.03}$  & - & $0.96^{+0.02}_{-0.02}$\\
$\Gamma$            &$1.643^{+0.001}_{-0.002}$&$1.57^{+0.06}_{-0.06}$. & $1.85^{+0.04}_{-0.05}$  &  $1.86^{+0.01}_{-0.01}$\\
$E_{\rm cut}$ [keV] &$115^{+1}_{-1} $         & $99^{+9}_{-8} $        & $500^{a}_{-41} $ & $500^{a}_{-8} $\\
Norm$_{\rm pl}$     &$2.2639^{+0.0002}_{-0.0001}$& $1.76^{+0.03}_{-0.03}$ & $1.9^{+0.7}_{-0.6}$ & $1.144^{+0.09}_{-0.08}$\\
   \hline
  \hline
\texttt{relxill} &  &  & & \\

  \hline
Index1        & - & - & $8^{+b}_{-2}$            & $0.935^{+0.4}_{-1.3}$  \\
Incl    [deg]      & - & - & $66^{+4}_{-7}$      & unconstrained       \\
$\log$[$\xi/$erg\,cm\,s$^{-1}$]    & - & - & $3.6^{+0.1}_{-0.1}$   & $3.5^{+0.1}_{-0.1}$    \\
A$_{\rm Fe}$ [$\odot$] & - & - & $10^{c}_{-3}$     & $4.5^{+1}_{-1}$    \\
Norm$_{\rm refl}$         & - & - & $0.025^{+0.004}_{-0.003}$ & $1.144^{+0.09}_{-0.08}$\\
   \hline
  \hline
Flux and bkg &  &  & &\\

  \hline  
  Norm$_{\rm bkg}$  & $2.1^{+0.2}_{-0.2}\times10^{-4}$ & $1.8^{+0.2}_{-0.2}\times10^{-4}$ & $2.2^{+0.2}_{-0.2}\times10^{-4}$& $1.6^{+0.2}_{-0.2}\times10^{-4}$ \\
  Flux$_{30-120\,{\rm keV}}$ Nu [erg\,cm$^{-2}$\,s$^{-1}$]  & $1.25^{+0.04}_{-0.04}\times 10^{-8}$  & $1.20^{+0.04}_{-0.04}\times 10^{-8}$  & $1.19^{+0.03}_{-0.04}\times 10^{-8}$ & $1.28^{+0.02}_{-0.02}\times 10^{-8}$ \\
  Flux$_{30-120\,{\rm keV}}$ INT [erg\,cm$^{-2}$\,s$^{-1}$]  & $1.235^{+0.001}_{-0.001}\times 10^{-8}$ & $1.23^{+0.01}_{-0.01}\times 10^{-8}$ & - &$1.23^{+0.01}_{-0.01}\times 10^{-8}$\\
  \hline
$\chi^2$ / d.o.f.   &  10160/3274$^{d}$              & 292/268                &  876/752 & 1138 / 779\\

  \end{tabular}
	\begin{list}{}{}
    	\item[$^*$] The systematics are not accounted for by the Cash statistic, thus the statistical errors quoted are unrealistically small. 
    	\item[$^a$] The high energy cut-off is pegged to its upper limit (500~keV)
    	\item[$^b$] The upper limit of the emissivity index is 10
 		\item[$^c$] The iron abundance is pegged to its upper limit (10)
    	\item[$^d$] This is the value of the Cash fit statistic. We also report here the $\chi^2$  value as reference 2642/3274. 
	\end{list}
    \label{tab:best_fit_models}
 \end{table*}

We generate light curves and calculate hardness ratios from the stray light 
observations of Cygnus~X-1 to characterize the state of the source. 
We analyze energy spectra of simultaneous \nustar\ and \integral\ observations 
to check the stray light cross-calibration at high energies. 
Fig.~\ref{fig:lc_hardness} top panel shows \textit{MAXI} and 
\textit{Swift/BAT} light curves of Cygnus~X-1 around the time the stray light observation was performed (red dashed line). 
The bottom panel shows the hardness ratio calculated using two \textit{MAXI} bands (see y-axis label in the figure).  
Looking at the hardness ratio versus time, 
we infer that the source is caught during the transition from the soft to the hard state when it was observed by \nustar\ in stray light. 
The \textit{MAXI} and \textit{Swift/BAT} light curves show that the thermal component is still 
present in the flux energy spectrum. 
Since the stray light observation allows us to extend the energy range above the optics bandpass, 
we consider the \nustar\ spectrum up to 160~keV, fitting for the background with a constant model.
The value of the background is mainly determined by the 120-160~keV energy range.
Again, we consider the \integralib\ spectrum strictly simultaneous with the \nustar\ stray light observation 
in order to match the exposure of the two instruments.
Fig.~\ref{fig:light_curves} shows how the relative short timescale variability of the source is 
seen consistently by the two instruments. 
We performed several fits to the energy spectrum with different combinations of models.
All the parameter values of our best fit results are listed in Table~\ref{tab:best_fit_models} and Table~\ref{tab:discussion_models}.

We start by comparing the two instruments above 30~keV. 
In this case, we choose to perform the fit using both Cash (\citealt{Cash1979}) and 
$\chi^2$ statistic in \texttt{xspec} (\citealt{Arnaud1996}).
By definition, the Cash statistic accounts for the errors based on the counts in each energy bins. 
The spectrum of \integralib\ has more than 8.5 million counts whereas the 
\nustar\ stray light spectrum has ~5300 counts with the same exposure of 40~ks. 
Therefore, the \integralib\ spectrum dominates the fit when we use the Cash statistics.
With the $\chi^2$ statistic, we can add a systematic error to the \integralib\ data; with 1$\%$ systematics, as suggested in the \textit{IBIS Analysis User Manual},  the results change only slightly. 
Fig.~\ref{fig:pl_highEn} shows the spectra fitted with a phenomenological cut-off power-law (Model 1) using $\chi^2$ statistic. 
The two spectra do not show any significant deviation from the cut-off power-law model with $\Gamma = 1.57 \pm 0.06$ and
high energy cut-off at $99^{+9}_{-8}$~keV.
We note that the cross-calibration constant between the two instruments is $1.02\pm 0.03$. 
This is a remarkably good agreement between the \nustar\ and \integralib\ fluxes that are 
constrained to be consistent within the 90$\%$ confidence level. 
We obtain the same result when the Cash statistic (cstat in \texttt{xspec}) is applied. 
In Table~\ref{tab:best_fit_models} we report the model flux for each of the instruments in the overlapping energy range (30-120~keV). In the case of \nustar, the value of the flux refers to the source model without considering the background.

When we consider the \nustar\ spectrum at lower energies, down to $3$ keV (the lower limit of the nominal \nustar\ band), 
the absorbed cut-off power-law model cannot reproduce the shape of the spectrum.
The top panel of Fig.~\ref{fig:residuals_gamma_fixed} shows the residuals of Model~1 when we include channels below 30~keV for \nustar. 
The upturn at low energies cannot be accounted for by just adding a thermal component (e.g. \texttt{diskbb}) 
since the disk temperature reaches values above 1~keV which is hotter than has been observed  
in the soft state of Cygnus~X-1 \citep[e.g.][]{Tomsick2014}.
Moreover, accounting for the low energy excess with a thermal 
component reveals clear residuals around the iron line (6.4~keV) region suggesting that the reflection emission contributes to the spectrum. 
\begin{figure}
	\includegraphics[width=\columnwidth]{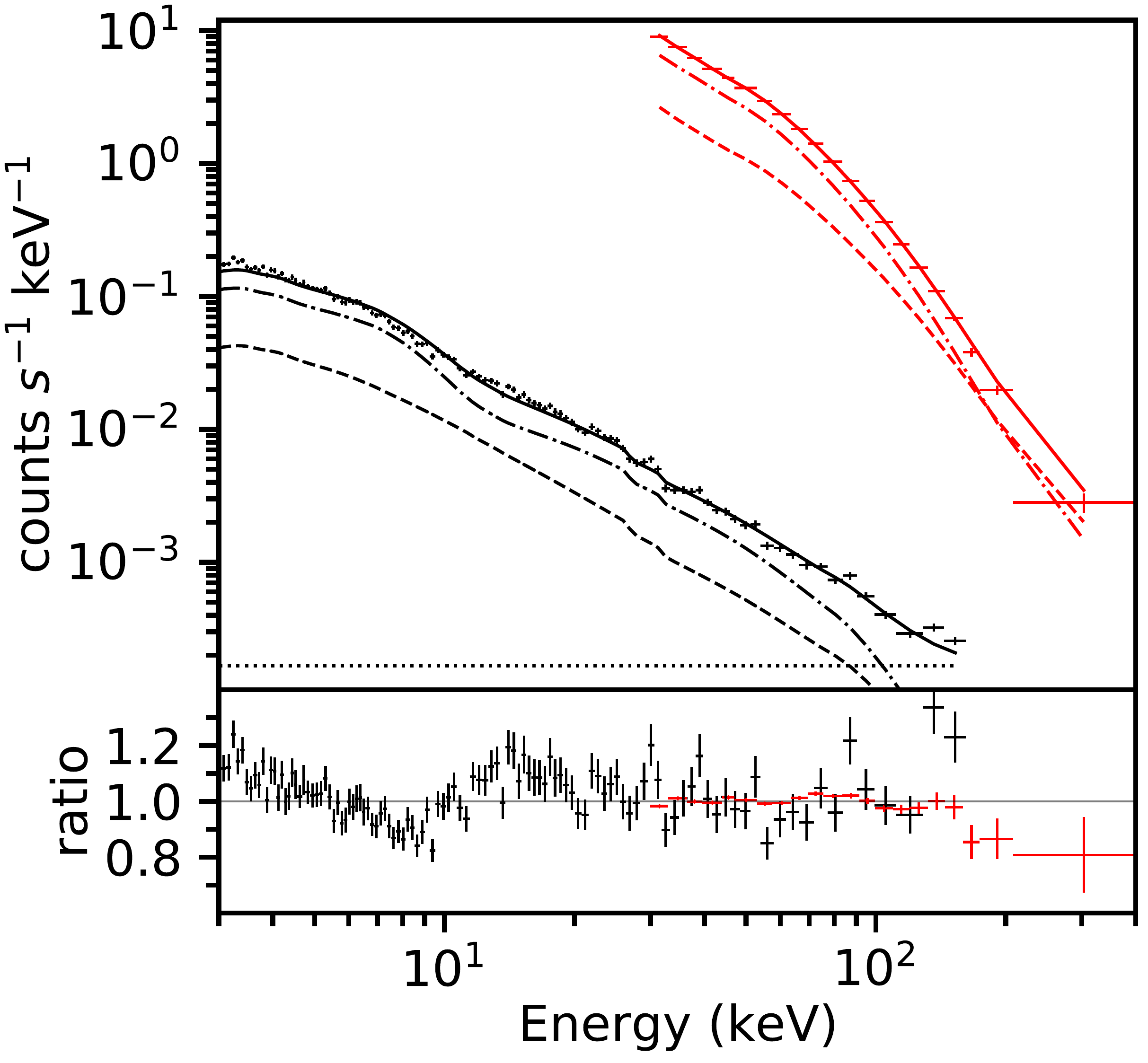}
    \caption{Best fit of the \nustar\ (black) and \integralib\ (red) spectra with Model~2b. Top panel: Flux energy spectrum in units of counts per seconds per keV, both data and model. The solid lines are the total models. The non-solid lines are the additive components of the models: the dotted line is the background model, the dashed lines are the continuum and the dashdotted lines are the reflection.  Bottom panel: residuals of the best fit.}
    \label{fig:tied_gamma_model}
\end{figure}

Since the stray light spectrum is well calibrated at low energies \citep[][]{Madsen2017a}, we proceed focusing on the \nustar\ stray light data. 
It is possible to fit the energy spectrum with 
a typical hard state model, Model~2: \texttt{TBabs*(cutoffpl + relxill)}. We fix the neutral column index at N$_{\rm H} = 6\times10^{21}$~cm$^{-2}$ as is typical for the source \citep[e.g.][]{Gilfanov2000,Tomsick2014}. Due to the relative high lower limit of the \nustar\ energy band (3~keV), it is not relevant to let the Hydrogen column density to be free in the fit. 
Fig.~\ref{fig:fit_only_nustar} shows that Model~2 leads to a good fit of the
\nustar\ energy spectrum. There are no broad structures in the residuals and  
the narrow residuals around 30-40~keV are internal background lines (see \texttt{nuskybkg} \citealt{Wik2014}). 
The parameters are sensible 
for the soft-to-hard state transition of Cygnus~X-1 (see Table~\ref{tab:best_fit_models}). 

Once we obtained an acceptable fit of the stray light spectrum with Model~2, 
we re-considered the \integralib\ spectrum. The bottom panel of Fig.~\ref{fig:residuals_gamma_fixed} shows the residuals of Model~2 when we only account for the different cross-calibration of the \integralib\ spectrum, but  without re-fitting the model. It is clear that the curvature of Model~2 is incompatible with the \integralib\ spectral shape. Thus we discard Model~2.

We fit Model~2 simultaneously to both the spectra with the continuum (cut-off power-law)
and reflection parameters free to vary (we refer to this fit as Model~2b). 
The parameter values are listed in last column of the Table~\ref{tab:best_fit_models}. 
Even though the fluxes of the two instruments are still compatible, 
the fit is not statistically good, $\chi^2$/d.o.f. = 1138/779. 
The high energy cut-off and the disk inclination are unconstrained, 
and the index of the emissivity profile (Index1) is extremely low. 
Moreover, Fig~\ref{fig:tied_gamma_model} shows that Model~2b 
presents some broad residuals in the \nustar\ spectrum around 
the Fe K$\alpha$ emission line (6.4~keV) and the Compton hump ($\sim20$~keV). 
Model~2b cannot reproduce the shape of the two spectra simultaneously 
and we argue the best fit is driven by the higher signal-to-noise of the
\integralib\ spectrum even with the addition of 1$\%$ systematic errors.
Despite of the fit result, Model~2b strengthens our confidence in 
the flux calibration of the stray light spectrum above 30~keV. 
The flux difference between the two instruments, in the overlapping band, is 
within 6$\%$ at 90$\%$ confidence level even for Model~2b. 
We also note that there is some level of degeneracy between the presumed source model and the background level when the source and the background spectral shapes are similar, which is likely affecting the result for Model 2b.

The poor fit of Model~2b might be cause by 
the wrong shape of the continuum model (power-law). 
It has been shown that the exponential cut-off of the 
\texttt{cutoffpl} model is not a good representation 
of the sharp drop of the data at high energies. 
However, when we considered a more physical model such as \texttt{nthComp + relxillCp},
the fit leads to the same conclusions, showing residuals very similar
to the ones shown in Fig~\ref{fig:tied_gamma_model}.




\begin{table}
\caption{Model~X is \texttt{TBabs*(diskbb + cutoffpl + gauss)} and Model~Y is \texttt{TBabs*(cutoffpl + relxill + (expabs*powerlaw))}. The column density is always fixed to $6\times10^{21}$~cm$^{-2}$. Statistical errors are quoted at $90\%$ confidence level.}
\begin{tabular}{ m{2.7cm}|m{2.2cm} | m{2.2cm}}
  \hline
Parameters  & 
Model X \footnotesize{($\chi^2$)} \newline \footnotesize{(disk + gauss)} & 
Model Y \footnotesize{($\chi^2$)} \newline \footnotesize{(high en tail)} \\
  \hline
Cross-cal            & $1.01^{+0.02}_{-0.02}$ & $1.00^{+0.02}_{-0.02}$  \\
$\Gamma$             & $1.56^{+0.04}_{-0.04}$ & $2.15^{+0.02}_{-0.01}$ \\
$E_{\rm cut}$ [keV]  & $99^{+6}_{-6} $ & $500^{a}_{-500} $ \\
Norm$_{\rm pl}$.     & $1.8^{+0.2}_{-0.2}$ & $6.04^{+0.1}_{-0.1}$  \\
T$_{\rm disk}$ [keV] & $1.0^{+0.1}_{-0.1}$   & - \\
Norm$_{\rm disk}$    & $466^{+234}_{-162}$   & - \\
  \hline
  \hline
relxill &  &   \\

  \hline
Index1                       & - & $9^{+1}_{-1}$ \\
Incl [deg]                   & - & $76^{+2}_{-2}$ \\
$\log$[$\xi/$erg\,cm\,s$^{-1}$] & - & $1.7^{+0.1}_{-0.1}$ \\
A$_{\rm Fe}$ [$\odot$]       & - & $10^{c}_{-0.4}$ \\
Norm$_{\rm refl}$            & - & $0.044^{+0.005}_{-0.002}$ \\
  \hline
  \hline
Gaussian &  &   \\
  \hline
LineE [keV]        &  6.4 (fixed)    &  - \\
Sigma  [keV]       &  $1.3^{+0.2}_{-0.3}$ &  - \\
Norm$_{\rm gauss}$ &  $0.07^{+0.02}_{-0.02}$ &  - \\

  \hline
  \hline
high en pl &  &  \\
  \hline
low E$_{\rm cut}$ [keV] &   &  100 (fixed) \\
$\Gamma$                    & - &  $2.5^{e}_{-0.03}$ \\
Norm                        & - &  $22^{+4}_{-2}$ \\
   \hline
 \hline
Flux and bkg &  & \\

  \hline  
  Norm$_{\rm bkg}$  & $1.8^{+0.2}_{-0.2}\times10^{-4}$ & $1.8^{+0.2}_{-0.2}\times10^{-4}$\\
  Flux$_{30-120\,{\rm keV}}$ Nu \newline [erg\,cm$^{-2}$\,s$^{-1}$]  & $1.22^{+0.03}_{-0.01}\times 10^{-8}$ &  $1.23^{+0.02}_{-0.04}\times 10^{-8}$  \\
  Flux$_{30-120\,{\rm keV}}$ INT \newline [erg\,cm$^{-2}$\,s$^{-1}$] & $1.23^{+0.01}_{-0.01}\times 10^{-8}$ & $1.23^{+0.01}_{-0.01}\times 10^{-8}$ \\
  \hline
$\chi^2$ / d.o.f.   & 850 / 780 & 944 / 777 \\

  \end{tabular}
	\begin{list}{}{}
    	\item[$^a$] The high energy cut-off is pegged to its upper limit (500~keV)
 		\item[$^c$] The iron abundance is pegged to its upper limit (10)
 		\item[$^e$] The slope of the high energy slope tail is pegged to the upper limit (2.5)
 		\end{list}
    \label{tab:discussion_models}
 \end{table}

\section{Discussion}
\label{sec:discussion}
We considered a stray light observation of Cygnus~X-1 simultaneous with \integral\ 
in order to test the calibration of \nustar\ at energies above 79~keV. 
In this Section we discuss the flux calibration of the stray light energy spectrum 
and the background at high energies. 
We also attempt to fit the broad band energy spectra with slightly 
more complicated models. 

\subsection{NuSTAR above 79~keV and background}
The stray light observations present a great opportunity to study X-ray sources above 79~keV with \nustar. It is important to demonstrate that the stray light energy spectrum
can be trusted above the nominal \nustar\ bandpass.
We found that the \nustar\ flux is well-calibrated and consistent with the \integral\ flux
at the 90$\%$ confidence level. 
This allows us to utilize the energy range when stray light observations are performed. 

Intentional stray light observations are performed by the \nustar\ team when a 
target is particularly bright in order to reduce the telemetry load. 
For example, the Crab was intentionally observed in stray light to calibrate the 
low energies \citep[][]{Madsen2017a}, and also bright black hole binaries 
such as MAXI~J1820+070 and MAXI~J1535-571. 
In those cases, if the signal-to-noise ratio is high enough, 
it is possible to analyze simultaneously 
the stray light spectrum together with the focused light spectrum 
to study the source emission above 79~keV. 
This type of analysis is possible even for serendipitous stray light observations, although the pointed observations will always have a better signal-to-noise ratio below 79~keV.

We caution the user about the background estimation, especially at high energies. 
We can reliably estimate the constant internal background by inspection of 
the flux level above 120 keV, as we explained in Section~\ref{sec:data_reduction_NuSTAR}. 
The astrophysical background is less trivial to evaluate.  
At high energies (above 20~keV), the stray light can penetrate the aperture stops 
and contaminates all the detectors (Fig.~\ref{fig:straylight_smoothing}). 
Therefore, it is not possible to extract the background spectrum from a region separated
from the stray light. 
At low energies (below 20~keV), the layers of aluminium (0.75 mm) and copper (0.13 mm) that form 
the aperture stops shield the photons. 
We computed the transmission spectrum ($T$) of the Al and Cu layers present in the aperture stop. 
We use $T(E) = e^{- \lambda\rho \alpha(E)}$ where $\lambda$ is the thickness of the aperture stops, $\rho$ and $\alpha$ are the density and the attenuation function\footnote{ www.physics.nist.gov/PhysRefData/Xcom/html/xcom1.html} of the element. 
At 20~keV the attenuation computed with coherent scattering is 3.4~g/cm$^{3}$ for Al and 33.8~g/cm$^{3}$ for Cu, thus, for each layer of Al and Cu the transmission at 20~keV is 10$^{-2}$, which quickly drops to 
10$^{-10}$ at 10~keV. 
Therefore, the `background' spectrum below 20~keV produced from a region outside the stray light 
region, which exceeds the constant internal background level, must be due to the Cosmic X-ray Background (CXB). 
We extracted the spectrum from a region different from, but close to, the stray light region highlighted in Fig.~\ref{fig:straylight_region} and found that below 20~keV the flux level of the CXB is only a few percent of the stray light flux during our Cygnus X-1 observation. 
We also fit the low energy spectrum with and without subtracting the CXB background 
and the result of the fit does not change. 
The CXB spectrum drops after 20~keV, thus the main source of background at high energies is the 
instrument internal background that we account for in our modelling. 
For fainter sources than Cygnus~X-1 we recommend using 
the \texttt{nuskybgd} model to estimate the 
background level of the observation more precisely.

\begin{figure}
	\includegraphics[width=\columnwidth]{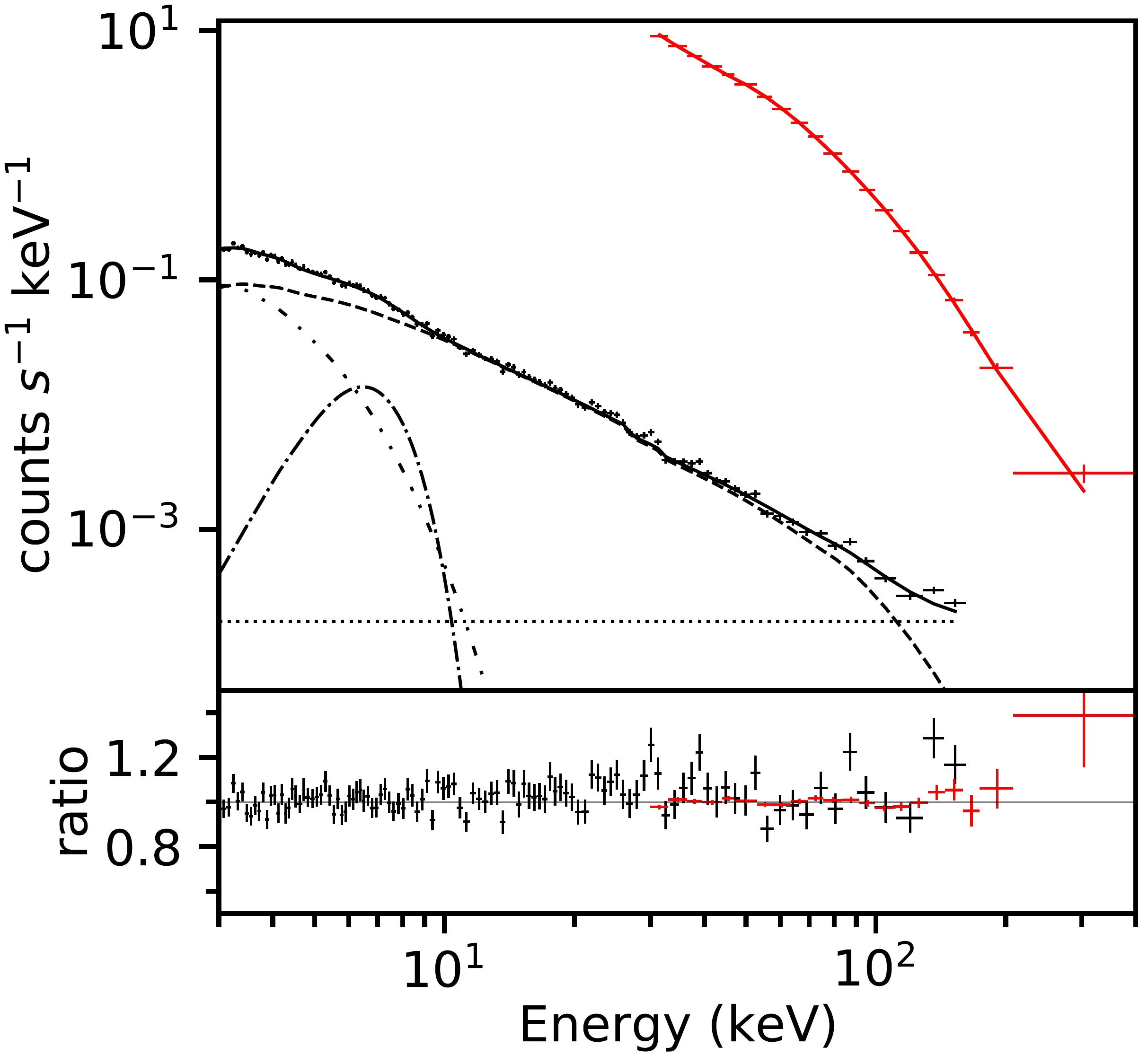}
    \caption{Best fit of the \nustar\ (black) and \integralib\ (red) spectra with Model~X. Top panel: Flux energy spectrum in units of counts per seconds per keV, both data and model. The solid lines are the total models. The non-solid lines are the additive components: the dotted line is the background model, the dashed line is the continuum, the dashdotted line is the Gaussian and the dashdotdotted line is the thermal disk.  Bottom panel: residuals of the best fit.}
    \label{fig:gauss_model}
\end{figure}

\begin{figure}
	\includegraphics[width=\columnwidth]{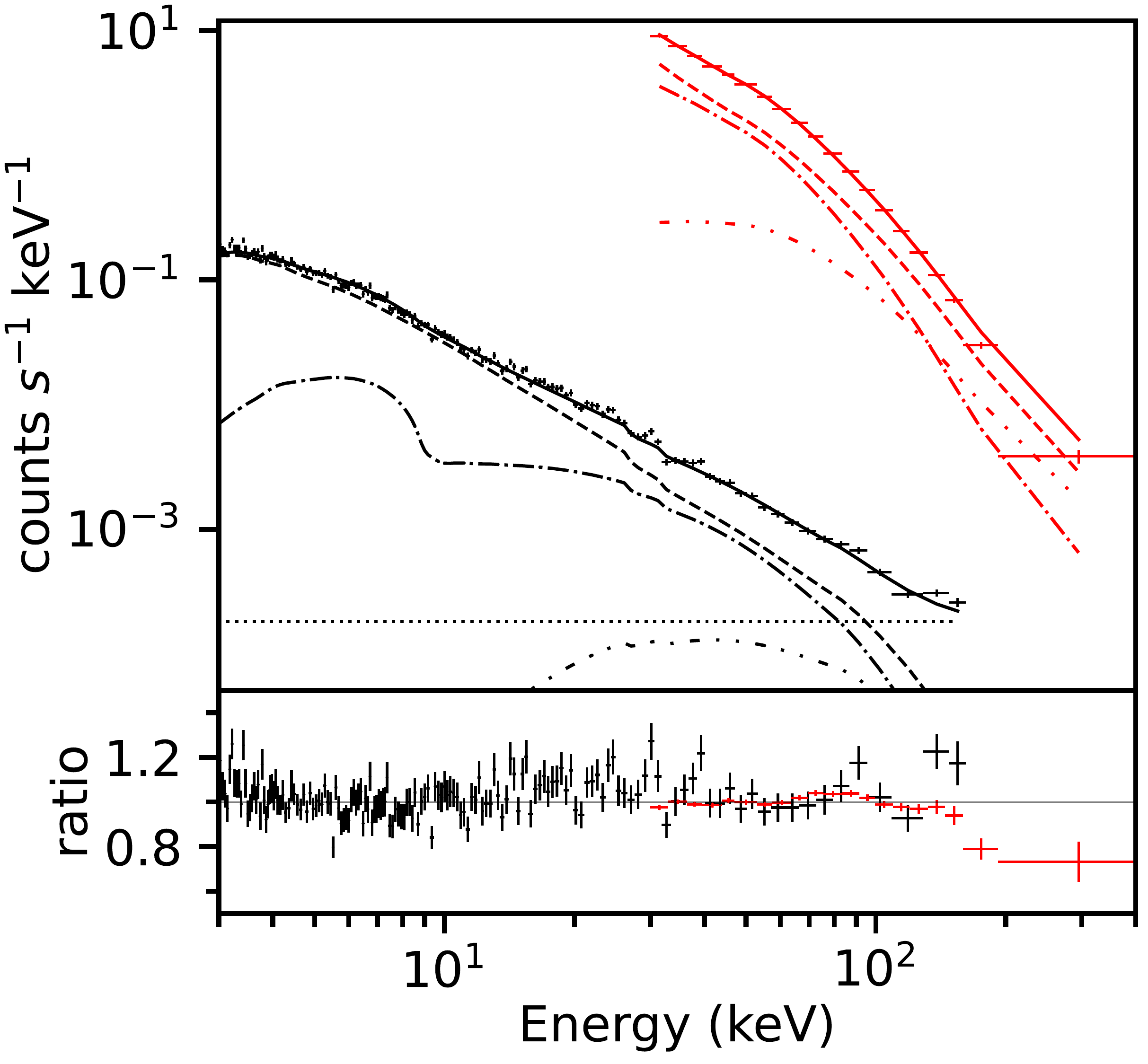}
    \caption{Best fit of the \nustar\ (black) and \integralib\ (red) spectra with Model~Y. Top panel: Flux energy spectrum in units of counts per seconds per keV, both data and model. The solid lines are the total models. The non-solid lines are the additive components: the dotted line is the background model, the dashed lines are the continuum, the dashdotted lines are the reflection and the dashdotdotted lines are the high energy tail. Bottom panel: residuals of the best fit.}
    \label{fig:highEn_model}
\end{figure}

\subsection{NuSTAR and INTEGRAL}

We found almost perfect agreement between the \nustar\ stray light and the \integralib\ flux in the
overlapping band 30-120~keV. We also argue that the spectral shape in this band is the same, since 
we can jointly fit both spectra with a high energy cut-off power-law. 
However, when we simultaneously fit the two spectra in the extended 3-400~keV energy range, 
Model~2b cannot reproduce the shape of the two spectra. 
We argue that since the source was caught during the state transition, the spectral analysis might require 
a more complicated model than the continuum + reflection addressed in our Model~2b. 
Here we propose two phenomenological test models that consider extra components contributing only either to the 
low or to the high energy band, but not to the shared energy range of the two instruments.

\paragraph{Thermal emission + simple reflection (Model X)} 
In this first case we consider the model 
\texttt{TBabs*(diskbb + cutoffpl + gauss)}. The model accounts for the soft excess in the \nustar\ spectrum 
with a strong thermal component, which is not required when we fit only \nustar. 
The reflection is simply a broad Gaussian line at 6.4~keV without Compton hump, 
thus there is no contribution to the \integralib\ energy band. 
Fig~\ref{fig:gauss_model} shows the best fit model and the residuals. 
The residuals do not show particular structures and the fit has an acceptable 
reduced $\chi^2$ of $850 / 755$. 
The second column in Table~\ref{tab:discussion_models} shows the parameter values of Model~X.
All the parameters are tied between the two instruments, apart 
from the cross-calibration constant. 
We note that this model reproduces the shape of the broadband spectrum 
by accounting for the soft-energy excess in the stray light using 
two additional components (thermal disk blackbody and iron line emission), 
neither of which contribute significantly at higher energies (above 20~keV).
However, we are inclined to reject this model because the temperature of 
the disk is too high to reasonably represent Cygnus~X-1 hard state,
and the lack of iron edge and Compton hump does not match our physical understanding of the reflection process.
\paragraph{High energy tail (Model~Y)} In this second case we consider 
a model component contributing at high energies in addition to Model~2b. 
This high-energy component has been detected and studied in Cygnus~X-1 since the source was observed at high energies ($>$100 keV) (e.g. \citealt{Ling1987,Phlips1996,McConnell2002,Bouchet2003,Pottschmidt2003,Zdziarski2012, Zdziarski2014, Cangemi2021}).
Our attempt is to understand if
neglecting this component in our model could artificially 
lead to a harder power-law spectrum, 
biasing our results at low energies when \integral\ and \nustar\ 
spectral slopes are tied together. 
We use the model \texttt{TBabs*(cutoffpl + relxill + expabs*powerlaw)}.
The last component is our simplistic attempt to account for the high energy tail (see \citealt{Connors2019}).
A more physical model is beyond the scope of this work. For a complete analysis of this subject, we refer to \citet{Cangemi2021}, which is one of the most recent attempt to fit the Cygnus~X-1 spectrum with Comptonization emitted by a hybrid population of thermal and non-thermal electrons (but see also the references previously mentioned in this paragraph). 
Fig~\ref{fig:highEn_model} shows the best fit model and the residuals performed with Model~Y. 
The high-energy tail component contributes to both the instruments and it considerably softens the 
power-law index ($\Gamma = 2.15^{+0.02}_{-0.01}$). 
We fixed the upper limit of the high energy power-law index to 2.5 and the lower limit 
of the exponential low energy cut-off to 100~keV, 
since without these constraints the additional component tends to compensate 
for the curvature introduced by the Compton hump in the \integralib\ spectrum
at relatively low energies. 
This would lead the high energy component to contribute to the 30-40~keV energy range, and below it, which is 
not its physical purpose. 
The fit prefers a much lower ionisation of the disk $\log\xi = 1.7 \pm 0.1$ compared to the previous models. 
Despite the fact that Model~Y should be considered a more physical model than Model~2b, 
the fit is statistically unacceptable with a reduced $\chi^2$ value of $944/777$.
Moreover, the \integralib\ residuals show that the new component overestimates the high energy flux. 
For all these reasons we tend to discard Model~Y.

We also tried to fit the high energy tail with the model \texttt{expabs*cutoffpl} with $\Gamma$ fixed at 2.5 and the high energy cut-off tied to the continuum component.
This model significantly improves the chi squared, 913 for 777 d.o.f., but it remains too high to accept this fit. Moreover, this fit does not require the continuum component which has a normalization compatible with zero. This is clearly non physical. If we remove the reflection component the fit shows significant residuals around the iron line.

Unfortunately, these tests seem to be inconclusive to establish the model that can fit the broadband spectrum of the two instruments. 
We think there is more complexity in the spectral shape of Cygnus~X-1, especially when such a broadband energy range is considered, and we postpone a more extensive analysis when pointed and stray light observations are performed simultaneously with \integral\ observations.

As a final remark we cannot completely rule out the possibility that there is 
cross-calibration slope difference between the two instruments since we did not find an acceptable model to fit the broad band \nustar+\integral\ spectrum. 
We note that the best measurement of the Crab spectrum using \nustar\ stray light \citep{Madsen2017b} produces a power-law slope from the Crab of $\Gamma= 2.106 \pm 0.006$. 
This is consistent with the spectral shape as reported by calibration measurement using \integralib, which reports a slope of $2.13 \pm 0.03$ \citep{Jourdain2008}, though there is some evidence for a spectral break above 100~keV measured by \integralib. 
The \nustar\ stray light observations of the Crab are background-dominated at these energies. 
However, the fact that the two instruments provide similar power-law indices indicates that a relative slope offset between the two is unlikely.

The spectral index for Model~2b ($1.85^{+0.04}_{-0.05}$), which considers 
only \nustar\ spectrum, is significantly softer than for Model~1 ($1.57\pm0.06$). This is due to the lack of high energy curvature in Model~2b ($E_{\rm cut} \sim 500$ keV). 
This arises from a mis-match in the statistical power of the \integralib\ data (even with added systematic error) and the \nustar\ data at low energies and is an example of the complexity of fitting data across such a wide bandpass.
We argue that the curvature of the model introduced 
by the reflection at high energies might be either incorrect for this source or might
require additional complexity due to the state of the source. 
We stress the importance of further investigations on comparing 
\nustar\ and \integral\ energy spectrum especially at high energies in other sources. 
This analysis will be the subject of a future work.

\section{Conclusions}
We analyzed the stray light flux energy spectrum of Cygnus~X-1, 
extending the \nustar\ energy range to 3-120~keV. 
We compare the stray light flux at high energies with the flux of 
the \integralib\ spectrum extracted during a simultaneous observation. 
We described how to account for the internal background 
which is the only source of background at high energies (above 30~keV).
We demonstrated that the astrophysical background can be neglected at low energies 
since it is only a few percents of the source flux. 
Based on the fact that the instrument produces the same spectral shape in the overlapping energy bands and produces compatible fluxes in the overlapping energy bands, we conclude that the \nustar\ stray light data can be used for scientific purposes up to 120 keV.
\newpage
\section*{Acknowledgements}
G. M. would like to thank Lorenzo Natalucci for the helpful comments. J.A.G. acknowledges support from NASA grant 80NSSC19K1020 and from the Alexander von Humboldt Foundation. The material is based upon work supported by NASA under award number 80GSFC21M0002. This work was supported by NASA under grant No. 80NSSC19K1023 issued through the NNH18ZDA001N Astrophysics Data Analysis Program (ADAP). This research has made use of the NuSTAR Data Analysis Software (NuSTARDAS) jointly developed by the ASI Science Data Center (ASDC, Italy) and the California Institute of Technology (USA).

\bibliography{library2022}{}
\bibliographystyle{aasjournal}



\end{document}